\documentclass[aps,prl,twocolumn,superscriptaddress,10pt]{revtex4-2}
\usepackage{setspace}
\usepackage{graphicx}
\usepackage{dcolumn}
\usepackage{bm}
\usepackage[colorlinks=true, linkcolor=blue, citecolor=blue, urlcolor=blue]{hyperref}
\usepackage{xcolor}
\usepackage{ulem}
\usepackage{soul}
\usepackage[utf8]{inputenc}
\begin{document}

\preprint{APS/123-QED}

\title{Origin of Glass-like Thermal Conductivity in Crystalline TlAgTe}\

\author{Shantanu Semwal}
\affiliation{
Department of Physics, Indian Institute of Technology Kanpur, Kanpur, UP 208016, India\\}%
\author{Yi Xia}
\affiliation{
 Department of Mechanical and Materials Engineering, Portland State University, Portland, 97201, OR, USA\\
}
\author{Chris Wolverton}
\affiliation{
 Department of Materials Science and Engineering, Northwestern University, Evanston, Illinois 60208, United States\\
}
\author{Koushik Pal}
\email{koushik@iitk.ac.in}
\affiliation{
Department of Physics, Indian Institute of Technology Kanpur, Kanpur, UP 208016, India\\}

\date{\today}

\begin{abstract}
Ordered crystalline compounds exhibiting ultralow and glass-like thermal conductivity are both fundamentally and technologically important, where phonon quasi-particles dominate their heat transport. Understanding the microscopic mechanisms that govern such unusual transport behavior is necessary to unravel the complex interplay of crystal structure, phonons, and collective excitations of these quasi-particles. Here, we use state-of-the-art first-principles calculations based on quantum density functional theory to investigate the origin of experimentally measured unusually low and glassy thermal conductivity in semiconducting TlAgTe that possesses disconnected chains of Tl atoms within its three-dimensional crystalline framework made up of distorted  AgTe$_4$ tetrahedra. Utilizing a unifying framework of anharmonic lattice dynamics theory that combine phonon self-energy induced  frequency renormalization, particle-like Peierls ($\kappa_l^P$) and wave-like coherent ($\kappa_l^C$) thermal transport contributions including three and four-phonon scattering channels, we successfully explain the experimental results both in terms of magnitude and temperature dependence. Our analysis reveals that TlAgTe exhibits several localized phonon modes arising from concerted rattling-like vibration of Tl atoms, which show strong temperature dependence and enhanced four-phonon scattering rates that are dominated  by Umklapp processes, suppressing $\kappa_l^P$ to ultralow values. The ensuing strong anharmonicity induced by local structural distortions, lone-pair electrons, and rattling-like vibrations of the heavy cations lead to a transition from particle-like behavior to wave-like tunneling characteristics of the phonon modes above 40 cm$^{-1}$, contributing significantly to $\kappa_l^C$ which increases with temperature.  Our analysis uncovers important structure-property relationship, which may be used in designing of novel materials with tunable thermal conductivity. 
\end{abstract}

\maketitle

Anharmonic crystalline semiconductors with ultralow lattice thermal conductivity ($\kappa_l$) are essential for applications in thermoelectrics, thermal barrier coatings, and data storage devices \cite{bell2008cooling,sootsman2009new} due to their unique heat transport properties. Strong anharmonicity causes atomic oscillators in crystalline materials to deviate significantly from simple harmonic motion, leading to enhanced phonon-phonon interactions that affect phonon quasi-particles and suppress $\kappa_l$. Therefore, understanding the microscopic mechanisms of thermal transport governed by phonons in the presence of strong anharmonicity is crucial for uncovering the underlying physics of phonons, designing new materials, and developing thermal management devices with enhanced performance. Several Tl containing compounds such as Tl$_3$VSe$_4$ \cite{xia2020particlelike}, TlInTe2 \cite{pal2021microscopic}, TlBiSe$_2$ \cite{ji2024influence}, TlAgSe \cite{pathak2024deciphering} have been shown to exhibit low thermal conductivity and high thermoelectric performance. TlAgTe which is an ordered crystalline semiconductor was shown experimentally to exhibit ultralow glass-like thermal conductivity \cite{kurosaki2007enhancement} and also predicted theoretically to show  high thermoelectric performance \cite{shafique2021multivalley}. However, a microscopic explanation of the underlying phonon physics that explains the glassy nature of $\kappa_l$ has remained elusive. While glass-like $\kappa_l$ is typically observed in amorphous \cite{allen1989thermal,sarkar2025glassy} and disordered materials \cite{allen1993thermal,simoncelli2025temperature}, it is rare in crystalline materials \cite{acharyya2022glassy,taneja2024high,rohj2024ultralow}, where low thermal conductivity may arise from the presence of rattler atoms, local structural distortions, or bonding heterogeneity  \cite{jana2016origin,qiu2014part,tadano2015impact,zhou2017promising,pal2018bonding, kim2015ultralow,Acharyya2022}, severely amplifying the lattice anharmonicity and inter-phonon interactions. As a consequence, the harmonic approximation (HA) of phonons may breakdown and phonon mean free path becomes comparable to the interatomic distances (Ioffe-Regel limit in space) and $\kappa_l$ approaches the amorphous limit  \cite{allen1989thermal,allen1993thermal,cahill1992lower}.  Recently, considerable efforts have been made  \cite{zhou2014lattice,tadano2015self,errea2014anharmonic,tadano2018quartic,errea2011anharmonic,Simoncelli2019,isaeva2019modeling,klarbring2020anharmonicity} to treat phonon-mediated phenomena in strongly anharmonic materials.

\begin{figure}[htbp]
\includegraphics[trim={6.6cm 0 5.5cm 0},clip=true,width=0.5\textwidth]{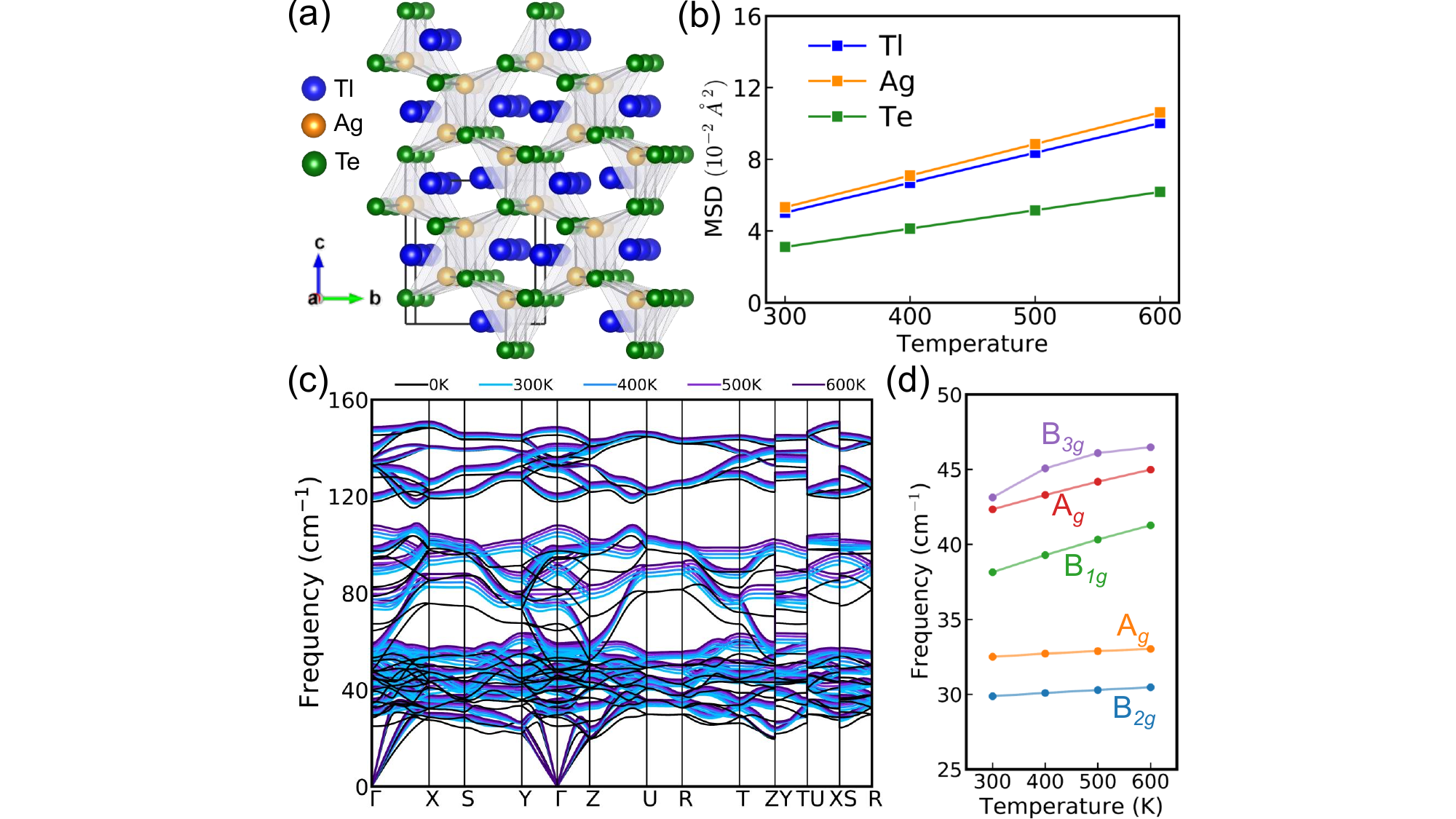} 
\caption{(a) Crystal structure of TlAgTe, where the distorted AgTe$_4$ tetrahedra (highlighted in grey) form the three-dimensional crystalline framework and Tl atoms are arranged as chains with the hollow space inside the structure. (b) Mean square displacement (MSD) parameters displaying large values for Tl and Ag. (c) Temperature-dependent anharmonically renormalized phonon dispersion of TlAgTe. (d) Raman modes are plotted as a function of temperature with their respective symmetry labels.}
\label{fig:scph}
\end{figure}

In this work, we investigate the origin of ultralow and glass-like $\kappa_l$ in TlAgTe combining a state-of-the-art high-order and unified anharmonic lattice dynamics framework using first-principles calculations based on density functional theory (DFT). TlAgTe possesses a three-dimensional crystalline framework consisting of distorted AgTe$_4$ tetrahedra [Fig. 1(a)] with chains of disconnected Tl atoms within its large hollow space and exhibits mixed covalent-ionic bonding with disparate bond strengths (see Fig. S1(a), Suppemental Material (SM) \cite{SI} for structural details). The calculated mean square displacement (MSD)  parameters of the atoms [Fig. 1(b)] show relatively large values for Tl and Ag atoms. Interestingly, despite the larger mass of Tl, its MSD is as large as that of the lighter Ag atoms, suggesting that Tl atoms are loosely bound and exhibit rattling-like motion within the crystal structure. Such large  MSD values are typical of cations in complex compounds such as Cu$_{12}$Sb$_4$S$_{13}$ \cite{xia2020microscopic}, AgTlI$_2$ \cite{zeng2024pushing}, TlInTe$_2$ \cite{jana2017intrinsic}, InTe \cite{zhang2021direct} exhibiting glassy or ultralow $\kappa_l$.  To get more insight into the bonding environment in this compound, we performed the projected electronic density of state calculations (SM \cite{SI} Fig. S1(b)) and electron localization function analysis (SM \cite{SI} Fig. S1(c)), which confirm the presence of localized  lone-pair electrons of Tl that is known to reduce $\kappa_{l}$ through augmenting the magnitudes of phonon scattering processes  \cite{jana2016origin,qiu2014part}. Therefore, we anticipate strong anharmonic effects arising from multiple factors in this material.

First, we analysed the vibrational properties of TlAgTe calculated within HA (i.e.,T = 0K) using the finite difference method implemented in Phonopy  \cite{togo2015first} (see SM  \cite{SI} for convergence tests, Fig. S2). We notice that phonon dispersion and phonon density of states (PhDOS) of TlAgTe can be clearly be divided into three groups (SM, Figs. S1(d,e)), which we denote as GR1: 30-55 cm$^{-1}$, GR2: 55-112 cm$^{-1}$ and GR3: 112-150 cm$^{-1}$. GR1 region is inhabited by multiple localized, nearly dispersion-less phonon branches originating mainly from the vibrations of Tl atoms, which is akin to rattling-like motion.  Since localized low-frequency modes in materials can strongly inhibit the propagation of phonons through the lattice by creating additional scattering channels \cite{pal2021accelerated,pal2021microscopic}, we expect that the dispersion-less phonon branches in TlAgTe would also have significant effects in lowering its $\kappa_l$. We provide visualizations of some of the low-frequency modes at $\Gamma$ point in SM \cite{SI} Fig. S1(f), which reveal concerted rattling of Tl atoms. 
Analysis of phonons under the HA neglects the anharmonic effects that might arise at finite temperatures due to phonon self-energy, which impacts phonon dispersions significantly in compounds like Tl$_3$VSe$_4$ \cite{xia2020particlelike} and TlInTe$_2$ \cite{pal2021microscopic}. It has been shown that strong quartic anharmonicity contributes to the self-energy of phonons significantly, shifting the bare phonon frequencies ($\omega_{\lambda}$) \cite{tadano2015self,errea2014anharmonic}, which is called phonon renormalization.

We used the self-consistent phonon (SCPH) method derived using the many-body Green's function theory \cite{tadano2015self} to iteratively calculate renormalized phonon frequencies ($\Omega_{\lambda}$) as a function of temperature (see SM \cite{SI} for details and convergence tests in Fig. S3(a)). We calculated the temperature-renormalized phonon dispersions [Fig. 1(c)], and PhDOS (SM \cite{SI} Figs. S3(b-c) \cite{SI}), which demonstrate strong effects of quartic anharmonicity on the phonon frequencies of TlAgTe, showing significant hardening of some low-energy phonon modes. For example, the first transverse optical mode changes from 25 cm$^{-1}$ (0K) to 30 cm$^{-1}$ (300K) at $\Gamma$ and 29 cm$^{-1}$ (0K) to 32 cm$^{-1}$ (300K) at S points. Phonon hardening is most pronounced in the GR2 region, followed by the GR1 region, where it is associated with vibrations involving Tl atoms. This intriguing property of hardening of phonon modes associated with the Tl atoms also resembles the rattling phonon modes observed in Tl$_3$VSe$_4$\cite{xia2020particlelike} and TlInTe$_2$ \cite{pal2021microscopic}, which too exhibit strong temperature dependence. Comparison of PhDOS  at 0K and 300K (see SM \cite{SI}, Figs. S3(b-c)) we see an increased localization at around 42 cm$^{-1}$ and 50 cm$^{-1}$ mainly arising from Tl, and around 95 cm$^{-1}$ mainly due to Te. This suggests that with an increase in temperature, the phonon branches in the GR1 and GR2 regions become more dispersion-less (localized) with larger phonon occupation. We present the temperature dependence of the Raman-active modes in Fig. 1(d) for future experimental validation. While our analysis focuses solely on renormalization from quartic anharmonicity, it is important to note that cubic anharmonicity can also induce a shift in phonon frequencies. However, as the cubic renormalization (bubble diagram) stems from second-order perturbation  \cite{maradudin1962scattering}, its impact is typically weaker than that of quartic corrections  \cite{tadano2015self,tadano2018quartic,xia2020particlelike}. 
\begin{figure*}[htbp]
    \centering
    \includegraphics[trim={3.3cm 0 3.8cm 0},clip=true,width=0.8\linewidth]{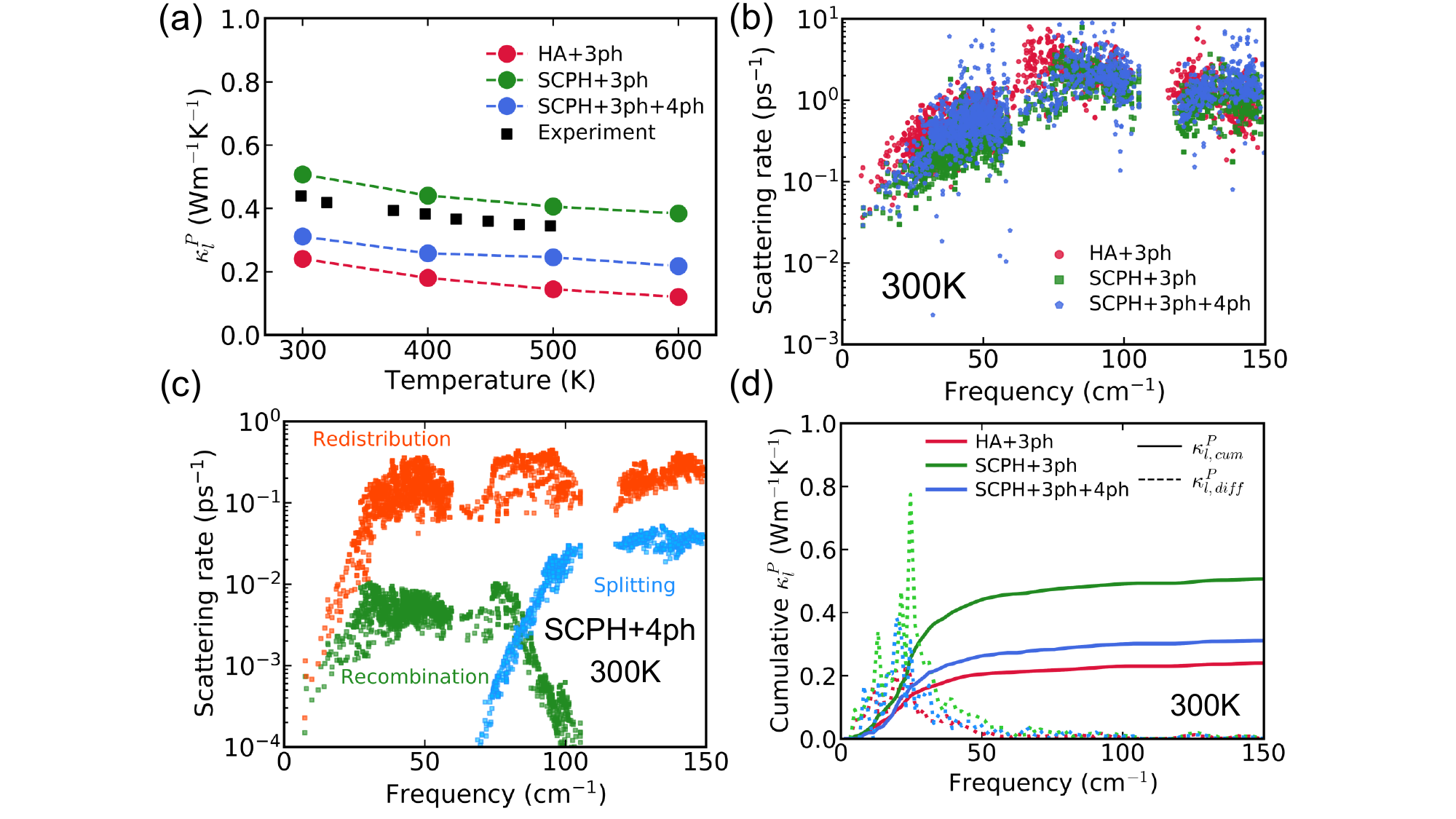}
    \caption{(a) Comparison of $\kappa^P_{l}$ under different levels of theory, where the black squares are the experimental result \cite{kurosaki2007enhancement}. (b) Comparison of three phonon (HA+3ph and SCPH+3ph) and three plus four phonon scattering (SCPH+3ph+4ph) rates. (c) Four-phonon (SCPH+4ph) Umklapp scattering rates for splitting ($\lambda \rightarrow \lambda'+\lambda''+\lambda'''+\mathbf{Q}$), redistribution ($\lambda +\lambda'\rightarrow \lambda''+\lambda'''+\mathbf{Q}$), and recombination ($\lambda +\lambda' +\lambda''\rightarrow \lambda'''+\mathbf{Q}$)  channels at 300 K, where $\lambda,\ \lambda', \ \lambda''$, and $\lambda'''$ are the phonon wave-vectors while $\mathbf{Q}$ is the reciprocal lattice vector. (d) Comparison of cumulative ($\kappa_{l,cum}^P$) and differential ($\kappa_{l,diff}^P$) values of $\kappa^P_{l}$ under different levels of theory. }
\end{figure*}

TlAgTe is a semiconductor with a calculated band gap of 0.45 eV (SM \cite{SI} Fig. S1(b)). Therefore, its thermal transport is primarily governed by phonons. First, we determine $\kappa_l$ using the Peierls-Boltzmann transport equation (PBTE), which has been very successful in describing the thermal transport properties of many crystalline solids and are widely used \cite{ShengBTE_2014,han2022fourphonon,pal2021microscopic,ward2009ab,broido2007intrinsic,omini1995iterative, omini1996beyond,chaput2013direct,fugallo2013ab,cepellotti2016thermal}. We used the ShengBTE code \cite{ShengBTE_2014} to calculate three-phonon scattering based on the full iterative solution to the PBTE, which gives particle-like contribution ($\kappa_l^P$) to the total $\kappa_l$, considering only the intraband phonon group velocity (see details and convergence tests in SM \cite{SI} Figs. S4(a-c)). In the first step of our analysis, we calculated $\kappa^P_l$ using HA and the three-phonon scatterings (3ph), henceforth referred to as (HA+3ph), as typically done in most studies.
 We find that the calculated $\kappa^P_{l}$(HA+3ph) is severely underestimated [Fig. 2(a)], which is only 55\% (0.24 $Wm^{-1}K^{-1}$) of the experimental value (0.44 $Wm^{-1}k^{-1}$) \cite{kurosaki2007enhancement} at 300K.  Moreover, the $T^{-1}$ dependence of $\kappa^P_{l}$(HA+3ph), generally obeyed by weakly anharmonic materials with dominant three-phonon scatterings, is far off from the weaker $T^{-0.48}$ dependence of the experimental data (see SM \cite{SI} Fig. S5(a)). Therefore, HA+3ph level of theory fails spectacularly for TlAgTe, which could be possibly due to the fact that temperature-dependent effects arising from strong anharmonicity were missing in the harmonic phonon frequencies and 3ph scattering rates.
 
Next, we utilize the renormalized frequencies ($\Omega_{\lambda}$) to determine phonon group velocity, heat capacity and 3ph scattering rates, resulting in elevation of $\kappa_l^P$ with a 110 \% increase to 0.51 $Wm^{-1}K^{-1}$ at 300 K when compared with HA+3ph values as shown in Fig. 2(a). We denote this level of theoretical correction as ``SCPH+3ph''. This increase in $\kappa^P_l$ is anticipated since the shift in phonon frequencies reduces the three-phonon scattering rates, as evidenced in Fig. 2(b). Since the phonon group velocities ($v_{\lambda}$) and molar specific heat ($C_v$) did not reveal any significant changes due to frequency renormalization (see SM \cite{SI} Fig. S6), the rise of $\kappa^P_{l}$(SCPH+3ph) $ \sim \sum_{\lambda} C_{\lambda}v^2_{\lambda}(\Omega_{\lambda})\tau_{\lambda}(\Omega_{\lambda})$ can be attributed to suppression of the three-phonon scattering rates and hence, an increase of phonon lifetime ($\tau_{\lambda}$). We fit the resulting $\kappa^P_l$(SCPH+3ph) with $T^{-\alpha}$ (with $\alpha>0$), that shows a weaker temperature dependence of $T^{-0.41}$ compared to the $T^{-1}$ decay found in the HA+3ph approximation, bringing temperature-dependence closer to experiment. However, the magnitude of $\kappa^P_l$ is largely overestimated due to the significant enhancement of phonon lifetime. Unlike the HA where the temperature dependence of $\kappa_l^P$ comes solely from the phonon occupation ($n_{\lambda}$)  \cite{maradudin1962scattering}, the weak temperature dependence of $\kappa^P_{l}$ at the SCPH level is influenced by $v_{\lambda}$, $C_\lambda$, and phonon scattering rates which  depended on $\Omega_{\lambda}$.
To understand the origin of the discrepancy above in the magnitude of $\kappa^P_l$, we realize that many strongly anharmonic crystalline compounds  show significantly large four-phonon scattering rates \cite{pal2021microscopic,xia2020particlelike,feng2017four,xia2020high}. Therefore, we determined the temperature-dependent four-phonon scattering for TlAgTe utilizing $\Omega_{\lambda}$'s, labelled as ``SCPH+4ph'', that shows that the magnitude of the four-phonon scattering rates are quite significant compared to the corresponding three-phonon scattering rates (see SM \cite{SI}, Fig. S7(a)). Therefore, it is essential to include the effects of both the temperature-renormalized frequencies along with the four-phonon scatterings. We also observe [Fig. S7(a)] that there is a larger number (however, lower in magnitude) of four-phonon scattering rates in the very low-frequency region ($<30$ $cm^{-1}$) due to increased combinations of phonon wave-vectors that satisfy energy and momentum conservation, leading to a larger four-phonon scattering phase space when compared to three-phonon processes. Hence, the inclusion of four-phonon scattering rates with the three-phonon processes enhances the scattering channels of phonons in this region. This level of theoretical correction is denoted by ``SCPH+3ph+4ph''. 

Further analysis (SM \cite{SI} Fig. S7(b)) reveals that four-phonon scattering is primarily due to the inelastic Umklapp (U) processes, which impede the organized flow of thermal energy and thus contribute to low thermal conductivity. The dominant U-scattering also justifies the relaxation time approximation (RTA) used in the FourPhonon code  \cite{han2022fourphonon} while calculating 4th-order scatterings (convergence test is provided in SM  \cite{SI} Fig. S4(d)). We further categorized the four-phonon processes into splitting, redistribution, and recombination channels [Fig. 2(c)] within the SCPH+3ph+4ph approximation at 300 K. Fig. 2(c) shows that the acoustic phonons ($<$30 $cm^{-1}$) scatter via the recombination and redistribution channel since the low-frequency mode cannot split further into lower energy phonons (splitting channel). Similarly, the high-frequency (100-150 cm$^{-1}$) modes cannot combine to give higher energy modes (recombination channel), so the scattering occurs significantly via the splitting (and redistribution) channel. The region between 30-100 cm$^{-1}$ is mainly dominated by scattering via the redistribution channel. The comparison of N-processes and U-processes for the 4ph scattering rates at 300 K (SM \cite{SI} Fig. S7(c)), reveals the dominant role of the U-scattering splitting, redistribution, and recombination processes. Inclusion of four-phonon scattering rates (SCPH+3ph+4ph) according to the Matthiessen's rule reduces phonon lifetimes, leading to a decrease in $\kappa^P_{l}$(SCPH+3ph+4ph), as shown in Fig. 2(a). $\kappa^P_{l}$(SCPH+3ph+4ph) at 300 K drops by a significant 37 \% to 0.32 $Wm^{-1}K^{-1}$ when compared to $\kappa^P_{l}$(SCPH+3ph).
\begin{figure*}[htbp]
\includegraphics[trim={3.5cm 0 3cm 0},clip=true, width=0.8\textwidth]{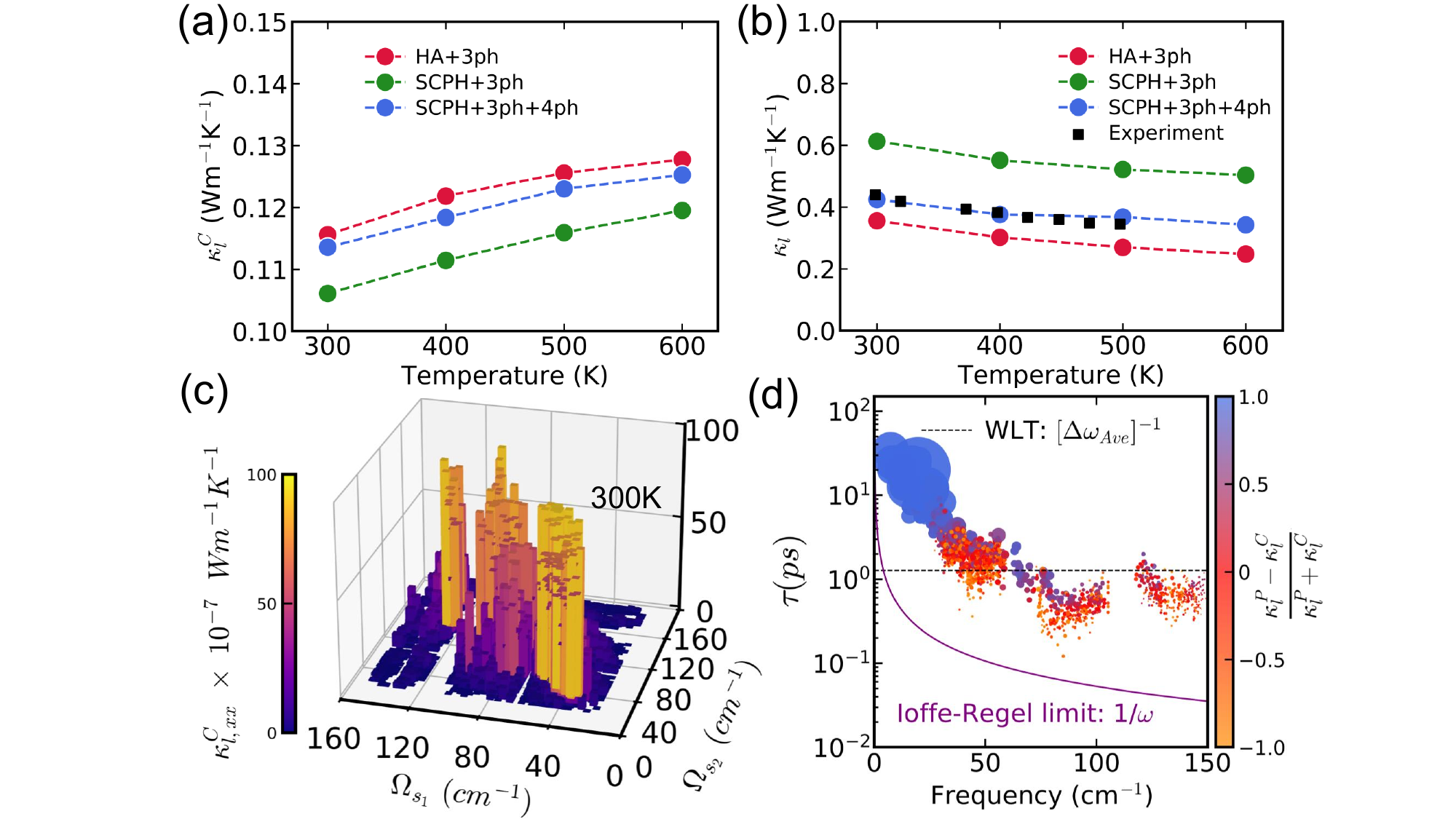}
\caption{(a) Comparison of coherent ($\kappa_{l}^{C}$) contribution to the lattice thermal conductivity for different levels of theory. (b) Comparison of total thermal conductivity ($\kappa_{l}=\kappa^P_{l}+\kappa_{l}^{C}$) for various levels of theory, compared to experimental result \cite{kurosaki2007enhancement}. (c) Contribution of coherent contribution as a function of renormalized phonon frequencies ($\Omega$) along the $xx$ direction ($\kappa^{C}_{l, xx}$), where subscripts $s_1$ and $s_2$ indicate two different phonon branches. (d) Phonon lifetime as a function of the phonon mode frequencies. The colors represent the contribution of phonon modes due to the particle-like and coherent transport mechanisms, quantified by $\frac{\kappa_l^P-\kappa_l^{C}}{\kappa_l^P+\kappa_l^{C}}$, where a value of 1 (blue) indicates pure particle-like propagation, a value of 0 (red) indicates equal contribution, and a value of -1 (orange) indicates pure coherent contributions. The black dashed line represents the Wigner limit in time (WLT) and the size of the data points represent relative magnitude of contribution to $\kappa_l$.}
\end{figure*}

To this end, we examine the mode-specific contributions to $\kappa^P_{l}$ by ploting cumulative $\kappa^P_{l}$ and the differential $\kappa^P_{l}$ [Fig. 2(d)], for all three levels of theory. The main contribution to $\kappa^P_{l}$ is from the acoustic ($<$30 $cm^{-1}$) and low-frequency optical phonon modes in the GR1 region (30 cm$^{-1}$ to 50 cm$^{-1}$), the cumulative plot for $\kappa^P_{l}$ continues to grow before saturating around 50 cm$^{-1}$ for all levels of theory. The differential $\kappa^P_{l}$ shows a slight shift of peaks in the low-frequency region when we compare HA+3ph with SCPH+3ph or SCPH+3ph+4ph due to hardening of phonon frequencies. Fitting the $\kappa^P_l$ data with $T^{-\alpha}$ shows a value of $\alpha=0.50$ (SM \cite{SI} Fig. S5(a)) for $\kappa^P_{l}$(SCPH+3ph+4ph) bringing the calculated temperature dependence of $\kappa_l^P$ closer to the experimental results ($\alpha=0.48$, SM \cite{SI} Fig. S5(a)). The origin of this behavior can be understood from the fact that $\tau^{-1}_{3ph} \sim n_{\lambda}\sim T$, $\tau^{-1}_{4ph} \sim n_{\lambda}^2\sim T^{2}$ and $\tau_{3ph+4ph}^{-1}\sim (aT+bT^2)$ dependence \cite{Feng2017}  on temperatures, where a and b are positive constants. Although the actual temperature dependence can deviate from this simple analysis, we still expect $\alpha_{3ph}<\alpha_{4ph}$. Consequently, we observe a slight increase in $\alpha$ for $\kappa^P_{l}$(SCPH+3ph+4ph) when compared to $\kappa^P_{l}$(SCPH+3ph) with $\alpha=$0.41. Although higher-order corrections to phonon frequencies and scattering rates give rise to weaker temperature dependence of $\kappa^P_l$ in accordance to experimental measurements, the magnitudes are still significantly off even with the highest level theory, i.e, SCPH+3ph+4ph. 

To reconcile the experimental result with our theoretical calculations, we realized that in our analysis of $\kappa_l^P$, we have so far considered the intraband phonon group velocity, leaving out the possibility of thermal energy transport due to the interband transition of phonons. According to the Allen-Feldman theory of thermal transport in glasses \cite{allen1993thermal}, thermal energy can be transported through tunneling between the localized phonon modes, which originates from the non-zero contribution from off-diagonal ($s\neq s_1$) elements of the phonon group velocity operator. In strongly anharmonic crystalline materials with local distortion, this wave-like transport mechanism (which gives rise to coherent contributions)  can coexist with the particle-like propagation. Simoncelli et al.  \cite{Simoncelli2019} recently developed a unified theory of thermal transport in crystals that bridges the particle-like phonon transport characteristic of ordered crystals with the wave-like (tunneling) behavior observed in glasses. We use this formalism to evaluate the wave-like transport contribution (denoted by $\kappa^{C}_{l}$), including contributions from both 3rd and 4th-order anharmonicity (see SM \cite{SI} for details). Fig. 3(a) shows $\kappa_l^C$ under different levels of theory, whose magnitude becomes significant when compared with $\kappa_l^P$. At 300 K, $\kappa_l^C$ is approximately 33\% of $\kappa_l^P$ for the SCPH+3ph+4ph case. Moreover, we see a rising coherent contribution with temperature, indicating a growing wave-like thermal transport of TlAgTe as the temperature increases. 

Combining $\kappa^{C}_{l}$ with $\kappa_l^P$ for the SCPH+3ph+4ph level of theory, we get the total thermal conductivity ($\kappa_{l}= \kappa_l^P + \kappa_{l}^{C}$) that agrees well with the experimental values [Fig. 3(b)]. We give $\kappa_{l}^P$, $\kappa_{l}^{C}$, and $\kappa_{l} $ for various levels of theory along different crystallographic directions (SM \cite{SI} Fig. S8) and their temperature-dependence (SM \cite{SI} Fig. S5). In Fig. 3(c), we present the mode-resolved plot of $\kappa_l^C$ (300 K) along the $xx$-direction, which shows large peaks in the GR1 and GR2 region indicating strong interband tunneling. This signifies that localized phonon modes in TlAgTe strongly contribute to the thermal transport via the wave-like mechanism, typical of glass-like materials. Unlike particle-like transport, in which only the low-lying modes mostly contribute, all phonon modes can contribute via the wave-like transport. The mode-resolved coherent contributions for the other directions are given in SM \cite{SI} Fig. S9. To get a deeper understanding of the characteristics of individual phonon mode, we plot their lifetime as a function of frequency [Fig. 3(d)]. The black dashed line represents the Wigner limit in time (WLT) \cite{simoncelli2022wigner}, which is the inverse of the average phonon interband spacing denoted by $[\Delta \omega_{Ave}]^{-1}=[\frac{\omega_{max}}{3N}]^{-1}$, where $\omega_{max}$ is maximum frequency in the phonon dispersion, and $N$ is the number of atoms in the primitive cell. The WLT has units of time and is a good estimate of a boundary around which phonon modes exhibit intraband particle-like transport ($\tau>[\Delta \omega_{Ave}]^{-1}$) to interband wave-like tunneling mechanism ($\tau<[\Delta \omega_{Ave}]^{-1}$). From Fig. 3(d), we see that the highly localized modes ($>$ 40 cm$^{-1}$) of TlAgTe contribute significantly via the wave-like mechanism. We provide mode-resolved lifetime for various levels of theory and the mean free path (MFP) plots in SM \cite{SI} Fig. S10. Our analysis shows that the scattering lifetime for all levels of theory is above the amorphous limit (Ioffe-Regel limit), which signifies that phonon wavevector remains well defined in the perturbative approach employed in our study. The MFP analysis also shows a similar cross-over region of the phonon modes from the particle-like transport to wave-like tunneling regimes, where WLT is replaced by the average bond length as a measure of the transport crossover boundary.

In summary, we used state-of-the-art first-principles calculations and a unified framework of anharmonic lattice dynamic theory that combines temperature-dependent phonon renormalization, high-order phonon scatterings, particle-like and coherent thermal transport contributions to unravel the origin of glass-like thermal conductivity in ordered crystalline semiconductor TlAgTe. Our analysis reveals that TlAgTe exhibits several localized phonon branches,  rattling-like vibrations, and strong anharmonicity. As a consequence, phonon frequencies show large temperature dependence and significantly high four-phonon scattering channels, dominated by the U-processes. The resulting $\kappa_l^P$ is significantly reduced and its temperature dependence is severely diminished, similar to those typically observed in disordered and glass-like compounds. Our results further reveal important role of the wave-like tunneling of phonons arising from the localized phonon modes above 40 cm$^{-1}$ that contribute quite significantly to $\kappa_l^C$. Hence, both particle-like and coherent transport mechanisms coexist in TlAgTe, whose combined contributions determined at the highest level of theory agree well with experimental results.

\begin{acknowledgments}
S.S. gratefully acknowledges financial support received from IIT Kanpur through an institutional fellowship. Y.X. acknowledges support from the US National Science Foundation through awards DMR-2317008 and CBET-2445361. K.P. thanks (i) IIT Kanpur for financial support through an Initiation Grant and (ii) the Anusandhan National Research Foundation for the PAIR grant (ANRF/PAIR/2025/000002/PAIR-B). We also acknowledge the use of computational resources provided by the HPC2013 cluster and the Param Sanganak supercomputing facility at IIT Kanpur.
\end{acknowledgments}
\bibliography{Bibilography}
\end{document}